\documentclass[twocolumn, aps, superscriptaddress, amsfonts,floatfix]{revtex4}

\usepackage{graphicx}
\usepackage{dcolumn}
\usepackage{bm}
\usepackage{color}
\usepackage{amssymb}
\usepackage{here}

\begin{document}

\preprint{APS/123-QED}

\title{Wing structure in the phase diagram of  the Ising Ferromagnet  URhGe close to its tricritical point investigated by angle-resolved  magnetization measurements}

\author{Shota~Nakamura}
\email{sna@issp.u-tokyo.ac.jp}
\affiliation{Institute for Solid State Physics, The University of Tokyo,  Kashiwa 277-8581, Japan}
\author{Toshiro~Sakakibara}
\affiliation{Institute for Solid State Physics, The University of Tokyo,  Kashiwa 277-8581, Japan}
\author{Yusei~Shimizu}
\affiliation{Institute for Solid State Physics, The University of Tokyo,  Kashiwa 277-8581, Japan}
\affiliation{Institute for Materials Research, Tohoku University, Oarai 311-1313, Japan}
\author{Shunichiro~Kittaka}
\affiliation{Institute for Solid State Physics, The University of Tokyo,  Kashiwa 277-8581, Japan}
\author{Yohei~Kono}
\affiliation{Institute for Solid State Physics, The University of Tokyo,  Kashiwa 277-8581, Japan}
\author{Yoshinori~Haga}
\affiliation{Japan Atomic Energy Agency (JAEA), Tokai 319-1106, Japan}
\author{Ji\v{r}\'{i}~Posp\'{i}\v{s}il}
\affiliation{Japan Atomic Energy Agency (JAEA), Tokai 319-1106, Japan}
\affiliation{Charles University in Prague, Faculty of Mathematics and Physics, DCMP, Ke Karlovu 5, 121 16 Prague 2, Czech Republic}

\author{Etsuji~Yamamoto}%
\affiliation{Japan Atomic Energy Agency (JAEA), Tokai 319-1106, Japan}

\date{\today}

\begin{abstract}
High-precision angle-resolved dc magnetization and magnetic torque studies were performed on a single-crystalline sample of URhGe, an orthorhombic Ising ferromagnet with the $c$ axis being the magnetization easy axis, in order to investigate the phase diagram around the ferromagnetic (FM) reorientation transition in a magnetic field near the $b$ axis.
We have clearly detected first-order transition  in both the magnetization and the magnetic torque at low temperatures, and determined detailed profiles of the wing structure of the three-dimensional $T$-$H_{b}$-$H_{c}$ phase diagram, where $H_{c}$ and $H_{b}$ denotes the field components along the $c$ and the $b$ axes, respectively.
The quantum wing critical points are  located at $\mu_0H_c\sim\pm$1.1~T and $\mu_0H_b\sim$13.5~T.
Two second-order transition lines at the boundaries of the wing planes rapidly tend to approach with each other with increasing temperature up to $\sim 3$~K. Just at the zero conjugate field ($H_c=0$), however, a signature of the first-order transition can still be seen in the field derivative of the magnetization at $\sim 4$~K, indicating that the tricritical point exists in a rather high temperature region above 4~K.
This feature of the wing plane structure is consistent with the theoretical expectation that three second-order transition lines merge tangentially at the triciritical point.

\end{abstract}

\pacs{Valid PACS appear here}
\maketitle

\section{Introduction}

The discoveries of uranium-based ferromagnetic (FM) superconductors, as represented by UGe$_2$ \cite{saxena2000superconductivity}, URhGe \cite{aoki2001coexistence}, and UCoGe \cite{huy2007superconductivity}, have had a great impact, because superconductivity and  ferromagnetism had been thought to compete with each other. 
 The superconducting properties in the above three uranium-based
 superconductors are extremely unusual, such as the microscopic coexistence of superconductivity and ferromagnetism  \cite{hattori2012superconductivity, ohta2010microscopic, PhysRevLett.102.167003, kotegawaJPSJ.74.705coexist}, 
possible occurrence of an odd-parity pairing \cite{saxena2000superconductivity, aoki2001coexistence, huy2007superconductivity}, 
the huge enhancement of $H_{\rm c2}$ exceeding the Pauli-limiting field in UCoGe \cite{huy2007superconductivity, 200915989}, and  re-entrant superconductivity (RSC) in URhGe \cite{levy2009coexistence}.
These anomalous behavior are observed around FM quantum phase transition, and hence 
 magnetic quantum fluctuations are considered to be responsible for the emergence of  such unusual superconducting states \cite{tokunaga2015reentrant, hattori2012superconductivity,taufour1742-6596-273-1-012017}.

We focus in this paper on the magnetic behavior of URhGe,
 which crystallizes in the orthorhombic TiNiSi structure with the space group $Pnma$ having a zig-zag chain of uranium atoms  along the $a$ axis \cite{TRAN199881}.
URhGe is known to be an itinerant ferromagnet  in which a magnetic moment  $M$ of $\sim 0.4$ $\mu_{\rm B}$/U aligns along the $c$ axis below $T_{\rm C} \sim 9.5$ K \cite{levy2005magnetic,hardy2011transverse}.
Magnetic anisotropy is very strong in URhGe, with the $c$ axis being the magnetization easy axis.
Owing to its strong anisotropy,  $T_{\rm C}$  of this compound can be tuned to zero by applying  a magnetic field $H$ along the $b$ axis, perpendicular to the spontaneous moment~\cite{hardy2011transverse}.
The situation is analogous to an Ising ferromagnet in a transverse magnetic field \cite{PFEUTY197079}, for which a quantum phase transition (QPT) accompanying a reorientation of the magnetic moment into a state with $M \parallel H$ can be expected at a finite critical field $H_{\rm R}$. 
 Previous studies by transport ($T\ge 0.5$~K), magnetic torque ($T\ge 0.1$~K) and magnetization ($T\ge 2$~K) measurements indicate that the transition occurs at $\mu_{0}$$H_{\rm R}\sim 12$ T in URhGe for $H \parallel b$ and becomes first order at low temperature \cite{levy2009coexistence,levy2005magnetic,hardy2011transverse,aoki2014superconductivity}.
Because the critical field $H_{\rm R}$ is very close to the field in which the RSC emerges, it has been argued that the magnetic fluctuations associated with the moment reorientation play an essential role of RSC~\cite{levy2005magnetic,levy2009coexistence,tokunaga2015reentrant}. 

Recently, a FM QPT in clean metals has attracted much interest because a first-order QPT is commonly observed~\cite{RevModPhys.88.025006}.
If $T_{\rm C}$ is decreased by some tuning parameter such as pressure, the nature of the transition changes from second order to first order  at a tricritical point TCP,  and by application of a small field parallel to the spontaneous moment surfaces or ``wings'' of first-order transition emerge~\cite{belitz1999first, PhysRevLett.94.247205}.
The edges of the wing planes are second-order transition lines, terminating at $T=0$ in quantum wing critical points (QWCPs)~\cite{PhysRevLett.115.020402}.
This type of ``$T$-$p$-$H$ phase diagram'' with pressure $p$ as a tuning parameter has been studied in itinerant FM compounds, such as UGe$\rm_2$ \cite{taufour2010tricritical, pfleiderer2002pressure, kotegawa2011evolution}, ZrZn$\rm_2$ \cite{uhlarz2004quantum}, URhAl \cite{PhysRevB.91.125115}, UCoGa \cite{mivsek2017pressure}, and an itinerant metamagnet UCoAl \cite{aoki2011ferromagnetic}.
In these systems, however, either high pressure (UGe$_2$, ZrZn$_2$, URhAl, UCoGa) or negative pressure (UCoAl) is required to tune $T_{\rm C}$ to zero, making it difficult to examine the magnetization behavior near QPT.
In contrast, as a magnetic field $H_b$ parallel to the $b$ axis being the tuning parameter, URhGe provides a good opportunity to investigate the whole FM phase diagram by various means.
Indeed, when a magnetic field is slightly tilted from the $b$ axis towards $c$ axis, the FM wing structure has been observed below TCP in the  ``$T$-$H_{b}$-$H_{c}$ phase diagram'' in URhGe~\cite{levy2007acute, levy2009coexistence}, where $H_{c}$ denotes the $c$-axis component of the magnetic field, conjugate to the order parameter (OP).
The location of TCP has been reported to be $\sim$ 2~K and  $>4$~K by thermoelectric power~\cite{gourgout2016collapse} and nuclear magnetic resonance (NMR)~\cite{kotegawa201573ge} experiments, respectively.
Remarkably, the zero resistivity region of the RSC at 50~mK exactly overlaps the wing QPT region in the $H_{b}$-$H_{c}$ plane~\cite{levy2007acute, levy2009coexistence}, further evidencing the close connection between RSC and the FM QPT.

It should be noticed that the RSC apparently emerges around the  first-order FM transition region. 
A possible explanation of this unusual behavior is that URhGe might be close to a quantum TCP~\cite{levy2007acute}.
There has been, however, a controversy regarding the location of TCP in URhGe~\cite{gourgout2016collapse,kotegawa201573ge}.
Further investigation is thus needed to clarify the FM QPT in URhGe.
Up to present, no direct magnetization measurement has been performed in URhGe in the QPT region; field variation of the magnetization in a $T=0$ limit has  been obtained by a $T^2$ extrapolation of the $M$ vs. $T$ data measured above 2 K for  $H \parallel b$~\cite{hardy2011transverse}. 
In the present paper, we have performed low-temperature angle-resolved dc magnetization measurements on URhGe in order to investigate the magnetization behavior near the wing structure of the FM QPT.
We obtained the $T$-$H_{b}$-$H_{c}$ phase diagram and determined the detailed profiles of the wing structure as well as the location of TCP.


\begin{figure}[t]
\begin{center}
\includegraphics[width=1\linewidth]{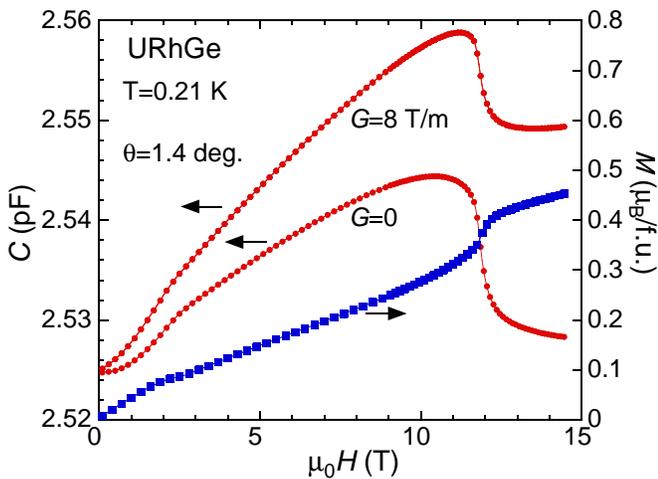}
\caption{ 
(color online). Example of the raw capacitance ($C$) data (solid circles) obtained at 0.21 K in magnetic fields tilted by 1.4$^\circ$  from  the $b$ axis in the $bc$ plane, with the field gradient $G = 0$ and 8  T/m. Taking a difference of these two yields the magnetization curve   (solid squares).
\label{Capa_G80_URhGe_0p2K_c}}
\end{center}
\end{figure}

\section{Experimental Procedures}
\begin{figure}[t]
\begin{center}
\includegraphics[width=1\linewidth]{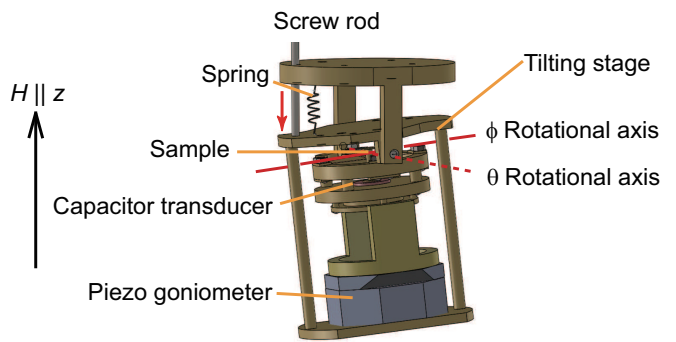}
\caption{ 
(color online). Schematic view of the two-axis rotation  device. $\theta$ and $\phi$ rotations are performed by a home-made tilting stage and  a piezo-stepper-driven goniometer, respectively. The angle of the tilting stage is controlled from the top of the insert by a screw rod that is thermally isolated.
\label{Cell_APS}}
\end{center}
\end{figure}


Single-crystalline URhGe was grown at JAEA, and cut into a rectangular shape with the 4.4 mg mass.
The present sample does not show superconductivity. It has been recognized that superconductivity as well as RSC only appears in stoichiometric samples of URhGe with a very small residual resistivity~\cite{doi:10.1143/JPSJ.77.094709}. 
 By contrast, the FM transition  is much more robust and does not change much even in doped systems  URh$_{0.9}$Co$_{0.1}$Ge and URh$_{x}$Ir$_{1-x}$Ge~\cite{tokunaga2015reentrant,PhysRevB.95.155138}. 
It should be noticed, however, that the sample quality might influence the magnetization behavior, in particular at the vicinity of QCPs.

DC magnetization  measurements were performed by means of a capacitively-detected Faraday magnetometer \cite{sakakibara1994faraday}.
 In this method, we detect a magnetic force ($M_{z}dH_z/dz$) proportional to the magnetization of the sample situated in an inhomogeneous field  as a capacitance change of a capacitive transducer. 
 Here, $z$ denotes the vertical axis, along which the magnetic fields up to 14.5~T were generated by a superconducting solenoid.
The capacitance transducer consists of a fixed plate and a mobile plate that is suspended by thin phosphor-bronz wires and can move  in proportion to an applied force.
We applied the field gradient of $G (=dH_z/dz) =8$~T/m in this experiment.
The sample was mounted on the capacitor transducer with varnish (GE7031) so that its $b$ axis is oriented close to the $z$ direction. 
In this situation, the $b$-axis component of the magnetization, $M_b$, is mainly detected. However, a huge magnetic torque component (${\bm M}\times {\bm H}$) is superposed on the output of the capacitor transducer due to the strong magnetic anisotropy. In order to eliminate the torque contribution, we measure the torque back ground with $G$ switched off ($G=0$), and subtract it from the data with $G$ switched on. 

Figure \ref{Capa_G80_URhGe_0p2K_c} shows an example of the data processing, in which the raw capacitance ($C$) data of URhGe obtained at 0.21 K in magnetic fields tilted by 1.4$^\circ$ from the $b$ axis in the $bc$ plane, are shown (solid circles) with two different field gradient values $G = 8$ T/m and $G = 0$.  The magnetization curve (solid squares) is obtained by taking a difference of the two data.
Further details of the data processing are given in Ref. \cite{sakakibara1994faraday}.
{Just at $\theta=0^\circ$, however, we had a difficulty in subtracting the torque background, as discussed later. 
The $G=0$ data is also useful to qualitatively estimate the $c$-axis component of the magnetization $M_c$, i.e., the OP of the FM state, under the fields near the $b$ axis.


 A $^3$He-$^4$He dilution refrigerator was used to cool the sample in the temperature range of 0.25 K $\leq T \leq$ 6 K.
The orientation of the URhGe crystal was precisely controlled in the $bc$ and $ab$ planes within an accuracy of less than 0.1$^\circ$ using a piezo-stepper-driven goniometer ($\phi$ rotation) combined with a home-made tilting stage ($\theta$ rotation)~\cite{PhysRevB.90.220502}, where $\phi$ ($-3^\circ \leq \phi \leq 3^\circ$) and $\theta$ ($-7^\circ \leq \theta \leq 7^\circ$) are the rotation angle in the $ab$ and the $bc$ planes, respectively.
Figure \ref{Cell_APS} shows a schematic view of the two-axis rotation device.
The two rotation axes, orthogonal to each other, intersect  the sample position. 
The angle of the tilting stage is varied by a screw, which is rotated from the top of the insert with a shaft that is thermally isolated.
The full details of the two-axis rotation device will be published elsewhere.
 In the present study, we measured the $\theta$ dependence of the magnetic responses in the $bc$ plane.


\section{Results}
\subsection{Torque component}
\begin{figure}[t]
\begin{flushright}
\includegraphics[width=0.95\linewidth]{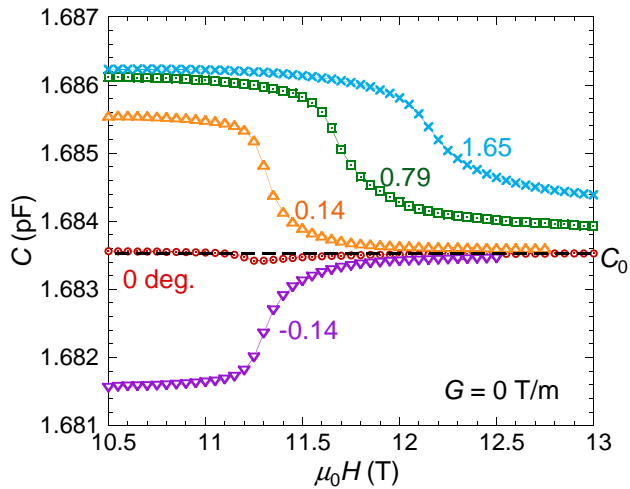}
\end{flushright}
\caption{(color online).  The raw capacitance data $C$ near $H_{\rm R}(\theta)$ with zero field gradient, measured at 0.5~K for several $\theta$ values, where $\theta$ is the angle between $H$ and the $b$ axis in the $bc$ plane. The black dotted line is zero torque state. These data were collected in a different run from the one in Fig.~\ref{Capa_G80_URhGe_0p2K_c}.
\label{deg_Torque-H_c}}
\end{figure}
\begin{figure}[t]

\begin{flushright}
\includegraphics[width=0.95\linewidth]{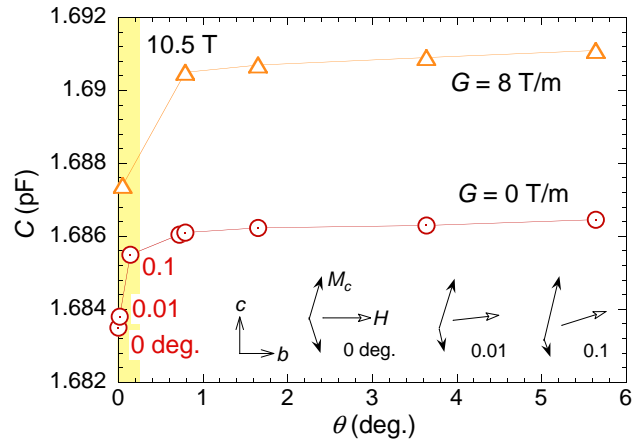}
\end{flushright}
\caption{(color online).  The angular $\theta$ dependence of the capacitance $C$ with two different field gradient values, $G = 0$ and 8  T/m, obtained at 0.5 K in 10.5 T below $H_{\rm R}$. In the yellow hatched region, the torque component changes so dramatically with $\theta$ that the precise evaluation of the magnetization becomes difficult. The inset schematically shows the $\theta$-evolution of the FM domains with positive and negative components of $M_c$ (solid arrows) in a magnetic field $H$ (open arrows). There is no bulk magnetization $M_c$ at $\theta=0^\circ$, and the zero-torque ($\tau=0$) state persists irrespective of the magnitude of $H_b$.
\label{Capa-deg_0p6K_c}}
\end{figure}

\begin{figure*}[t]
\begin{center}
\includegraphics[width=0.95\linewidth]{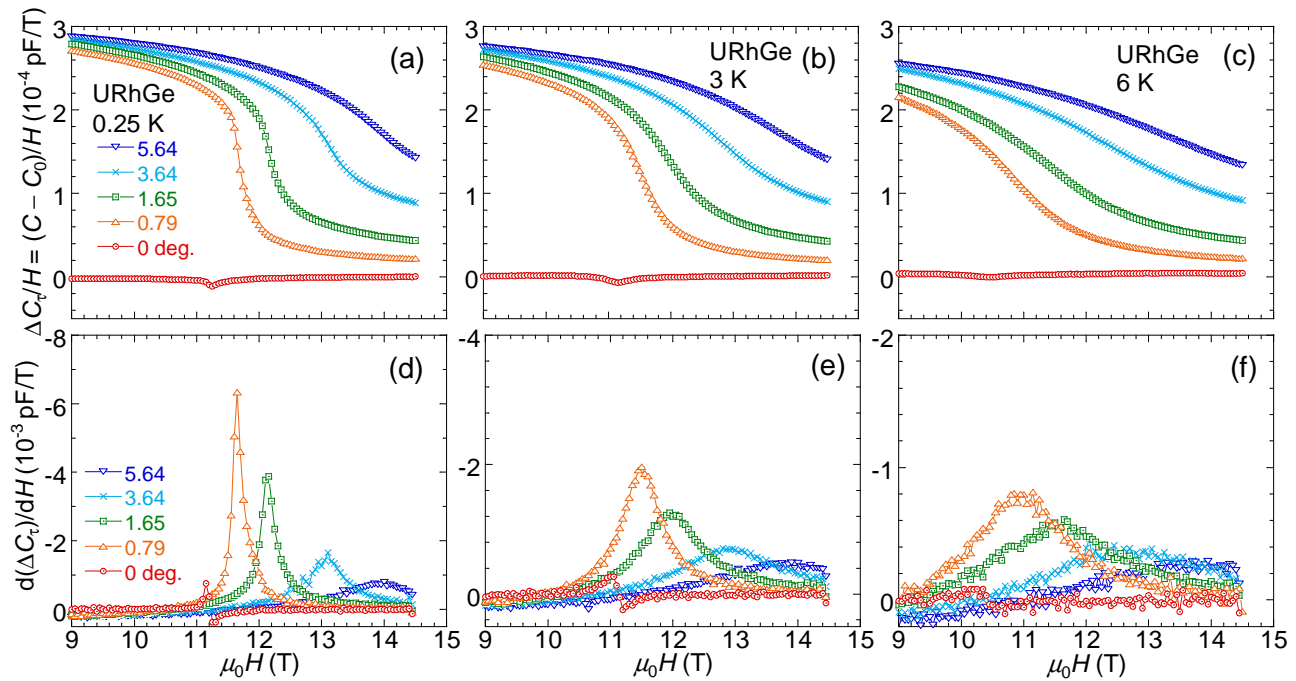}
\caption{(color online). Magnetic torque divided by field, $\Delta C_{\rm\tau}/H$, of URhGe measured at (a) 0.25, (b) 3, and  (c) 6 K in fields near $H_{\rm R}(\theta)$ with $\theta = 0$$^\circ$, 0.79$^\circ$, 1.65$^\circ$, 3.64$^\circ$, and 5.64$^\circ$,  together with the differential curves $d(\Delta$$C_{\rm \tau})$$/dH$ ((d)-(f)).
\label{Torque-theta-T_c}}
\end{center}
\end{figure*}
\begin{figure}[t]
\begin{flushright}
\includegraphics[width=0.9\linewidth]{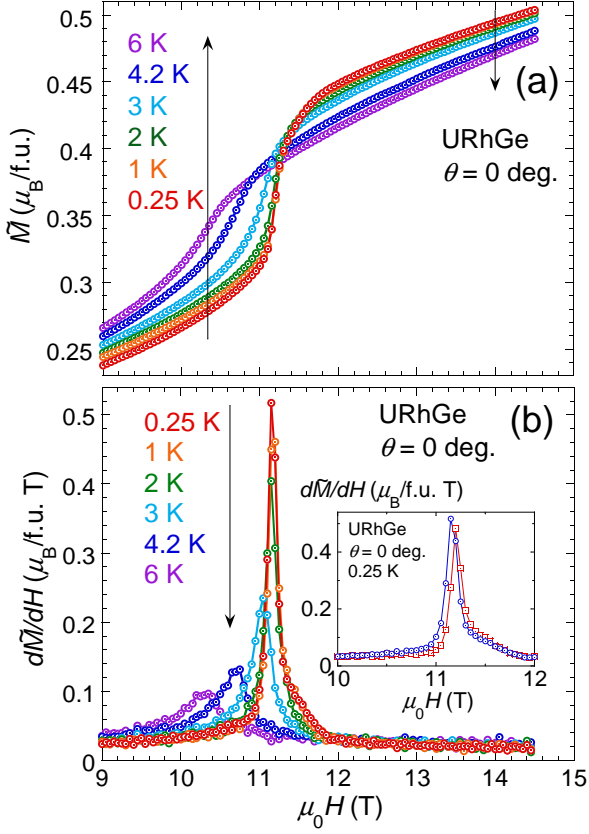}
\end{flushright}
\caption{(color online).  (a) $\tilde{M}(H)$  of URhGe measured in the magnetic field along the $b$ axis.
 (b) The field derivative $d\tilde{M}/dH$ of the magnetization curves in Fig. \ref{theta_zero_MH}(a). In these figures, only the down-sweep traces are plotted for simplicity. The inset of Fig.~\ref{theta_zero_MH}(b) shows  $d\tilde{M}/dH$ at $T=0.25$~K and $\theta=0^\circ$ for both up- and down-field sweeps.
\label{theta_zero_MH}}
\end{figure}
\begin{figure}[t]
\begin{flushright}
\includegraphics[width=0.9\linewidth]{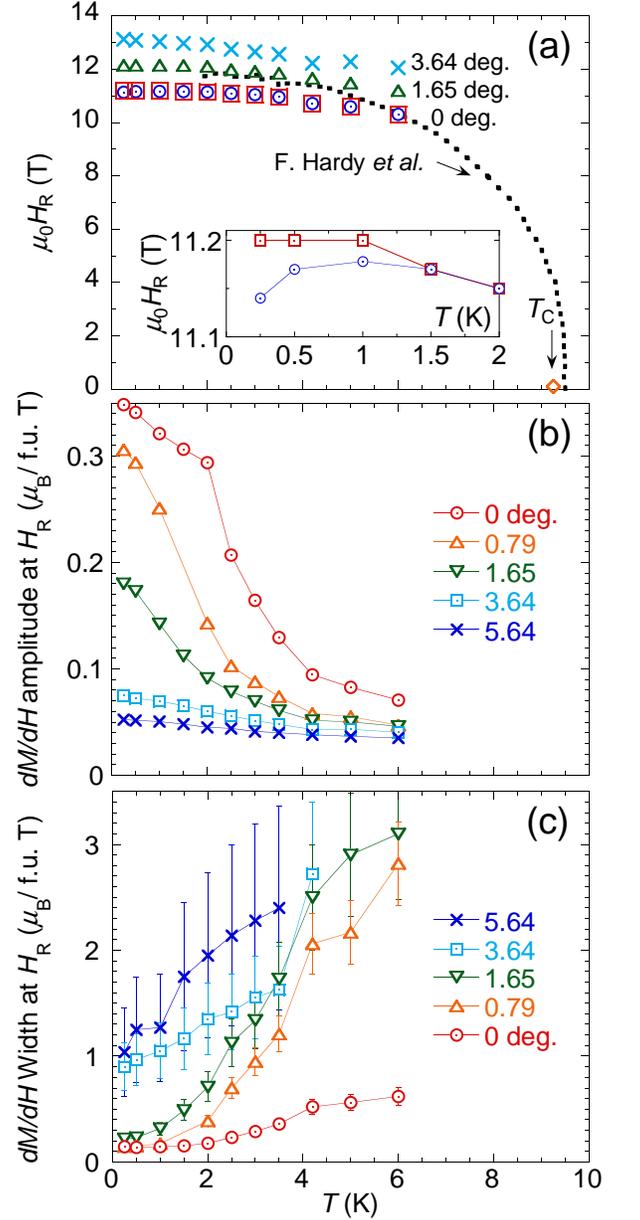}
\end{flushright}
\caption{(color online).  (a)  Temperature dependence of  $H_{\rm R}(\theta)$ obtained from the present measurements at $\theta = 0$$^\circ$, 1.65$^\circ$, and 3.64$^\circ$,
 together with the results of the previous study (dashed line)~\cite{hardy2011transverse}. The inset is an expanded plot for $\theta=0^\circ$, indicating $H_{\rm R}(T)$ defined at up-sweep (open squares) and down-sweep (open circles) fields.
 (b) Temperature evolution of the peak amplitude of $d\tilde{M}/dH$ at $H_{\rm R}$, obtained at several $\theta$. (c) Temperature evolution of the transition width, obtained at several $\theta$.
\label{theta_zero_TCP_HR}}
\end{figure}
\begin{figure*}[t]
\begin{center}
\includegraphics[width=0.95\linewidth]{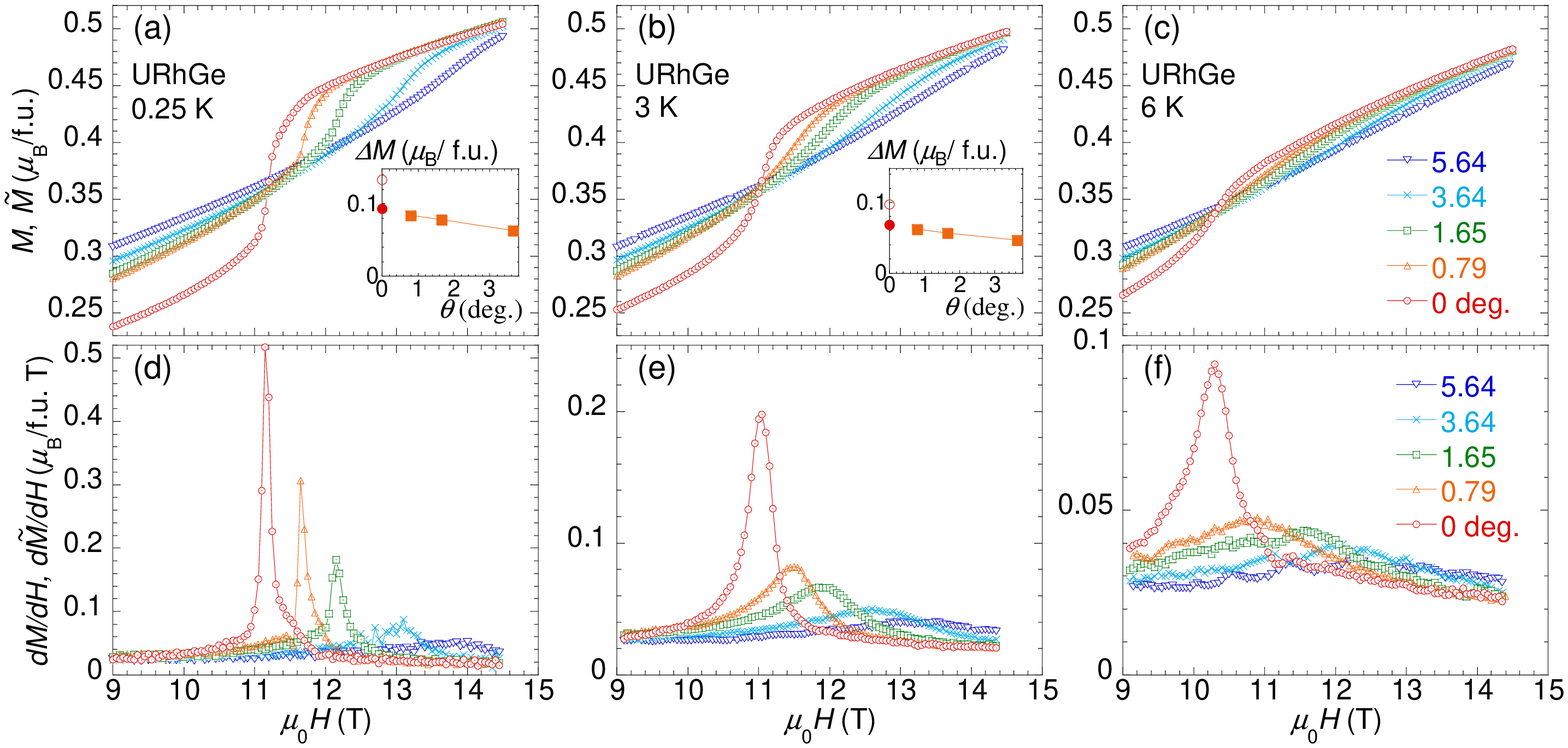}
\caption{(color online). The  magnetization curves $M(H)$ of URhGe near $H_{\rm R}(\theta)$ ($\theta = 0.79$$^\circ$, 1.65$^\circ$, 3.64$^\circ$, and 5.64$^\circ$), measured at (a) 0.25, (b) 3, and  (c) 6 K, together with their differential curves $dM/dH$ for (d) 0.25, (e) 3, and  (f) 6 K.
 For comparison, $\tilde{M}(H)$ and $d\tilde{M}/dH$ at $\theta = 0$$^\circ$ are also plotted. The insets in Figs. \ref{MH_deg_raw}(a) and \ref{MH_deg_raw} (b) show the angular variation of the magnetization jump $\Delta M$ ($\theta\ge 0.79^\circ$, solid squares) measured at 0.25 and 3 K, respectively. The solid circles are the linear extrapolation of  $\Delta M$ to $\theta=0$$^\circ$, which are a factor of 0.7 smaller than $\Delta \tilde{M}$ measured at $\theta=0$$^\circ$.
\label{MH_deg_raw}}
\end{center}
\end{figure*}

Figure \ref{deg_Torque-H_c} shows the raw capacitance ($C$) data near $H_{\rm R}$ with zero field gradient ($G = 0$) measured at $T=0.5$~K for several $\theta$ values, where $\theta$ is the angle between $H$ and the $b$ axis in the $bc$ plane. 
The dashed line at $C_0=1.683$~pF indicates the capacitance value at $H=0$. The capacitance difference $\Delta C_{\tau}$ $= C - C_{\rm 0}$ is thus proportional to the torque component $\tau = {\bm M} \times {\bm H}$.
There is a huge torque contribution below the reorientation field $\mu_0H_{\rm R}(\theta)\sim$ 12-13 T for $|\theta|\ge0.14^\circ$ coming from the large $M_c$ component.
Interestingly, the $G = 0$ data at $\theta = 0^\circ$, in which the magnetic field direction is precisely adjusted to the $b$ axis, show virtually no torque contribution.
This is because a perfect alignment of the magnetic field along the $b$ axis ($H_c=0$) yields an equal population of the FM  domains with $M_c$ pointing along $+c$ and $-c$ directions (the inset of Fig. \ref{Capa-deg_0p6K_c}), resulting in the zero-torque state even below $H_{\rm R}$.
The degree of the domain alignment changes with $\theta$, and saturates above 0.79$^\circ$. Note that the torque changes its sign for a negative $\theta$ value, as expected.
Above $H_{\rm R}$, the torque component almost vanishes for $\theta=0.14^\circ$, indicating that the  magnetic moment becomes almost parallel to the field direction. 
For  $\theta>0.79^\circ$, on the other hand, a finite torque remains even well above $H_{\rm R}(\theta)$. This is due to an intrinsic magnetic anisotropy of the system.
In Fig.~\ref{Capa-deg_0p6K_c},  we plot the $C$ value at $\mu_0H=10.5$~T and $T=0.5$~K with $G=0$ (open circles) and $G=8$~T/m (open triangles) for several $\theta$ values. The $G=0$ data represents the domain alignment as a function of the $c$-axis field $\mu_0H_c~[T]=10.5\sin\theta$. 
The data clearly show that the single domain state is reached at $\theta\sim 0.8^\circ$, or $\mu_0H_c\sim 0.15$~T. 


Figure \ref{Torque-theta-T_c} shows the field variation of $\Delta C_{\tau}/H$, the quantity proportional to $M_c$, measured at (a) 0.25, (b) 3, and  (c) 6 K in a field range near $H_{\rm R}$ for the angles $\theta = 0$$^\circ$, 0.79$^\circ$, 1.65$^\circ$, 3.64$^\circ$, and 5.64$^\circ$. 
The differential curves $d(\Delta$$C_{\tau})$$/dH$ are also shown in Figs. \ref{Torque-theta-T_c}(d)-\ref{Torque-theta-T_c}(f).
For $\theta = 0$$^\circ$, the domain state with zero magnetic torque persists up to $\mu_0H_{\rm R}(0^\circ )=11.2$~T, where a small kink appears upon the moment reorientation.
 For $\theta =0.79^\circ$ and $T=0.25$~K, the sudden collapse of $\Delta C_{\tau}/H$ seen at $H_{\rm R}(0.79^\circ)=11.7$~T    indicates a first-order transition. This transition becomes broader and shifts to the higher field side with increasing $\theta$ and decreasing $T$.
These features are more clearly seen in the differential data [Figs. \ref{Torque-theta-T_c}(d)-(f)].

These torque data thus directly probe the behavior of the OP across the transition, and can be used to construct the wing structure phase diagram for $\theta > 0.8$$^\circ$.
As mentioned above, however, the torque component is not  so sensitive to the phase transition very close to $\theta = 0$$^\circ$ because of the domain formation.
In order to explore the phase transition for $\theta \approx 0$$^\circ$, in particular the TCP, we evaluate the field variation of the magnetization  in the following.


\color{black}

\subsection{Magnetization}

 
Magnetization curves for various field angle $\theta$ near the $b$ axis can be obtained from the capacitance data with $G=8$~T/m by subtracting a torque background ($G=0$ data), on the basis of the assumption that the torque contribution is the same for $G=0$ and 8~T/m. In most cases, this condition holds with  good accuracy. As explained later, however, we found  it  difficult to fulfill this condition at $|\theta|\lesssim 0.1^\circ$ for a technical problem, and as a consequence some residual torque contribution remains in the magnetization curve at $H\lesssim H_{\rm R}$ for $\theta\approx 0^\circ$. For this reason, we denote the magnetization curve at $\theta=0^\circ$ by $\tilde{M}(H)$, and distinguish it from the $M(H)$ data for $\theta>0.1^\circ$ for which the torque component is properly subtracted.
 
Figures \ref{theta_zero_MH}(a) and (b) show $\tilde{M}(H)$ of URhGe and the differential curve $d\tilde{M}/dH$, respectively,  obtained at 0.25, 1, 2, 3, 4.2, and 6 K. 
At 0.25 K, a magnetization jump  is observed at $\mu_{\rm 0}$$H_{\rm R}(0)=11.2$ T.
$\tilde{M}(H)$  reaches $\sim$0.46 $\mu_{\rm B}$/U above $H_{\rm R}$, in agreement with the previous result~\cite{hardy2011transverse}. This magnetization value is very close to the spontaneous magnetization $M_c$ at $H=0$, in accordance with a simple picture of the moment reorientation
 from the easy $c$ axis to the $b$ axis at $H_{\rm R}$~\cite{tokunaga2015reentrant,levy2009coexistence,hardy2011transverse}. 
Just above $H_{\rm R}$, there is a small shoulder-like anomaly, which is also seen in $d\tilde{M}/dH$ as a small hump.
As shown in the inset of Fig.~\ref{theta_zero_MH}(b), a small hysteresis is observed in the transition at 0.25~K, implying the transition to be of first order.
With increasing temperature, the magnetization jump becomes broader and weaker, and the critical field shifts to the lower field side.  
This change of the transition behavior becomes prominent above 2~K. 
Surprisingly, however, the peak feature in $d\tilde{M}/dH$ can be seen even at 6 K.

 Figure \ref{theta_zero_TCP_HR}(a) shows temperature dependence of $H_{\rm R}(\theta,T)$ obtained from the present torque and magnetization measurements at $\theta = 0$$^\circ$, 1.65$^\circ$, and 3.64$^\circ$, together with the previous results (dotted line) \cite{aoki2011ferromagnetic}.
Here $H_{\rm R}(\theta,T)$ is determined by the position of the peak of $d(\Delta$$C_{\tau})$$/dH$ and $d\tilde{M}/dH$ in Figs. \ref{Torque-theta-T_c} and \ref{theta_zero_MH}.
The results at $\theta=0$ is qualitatively the same as the previous reports, and the line of $H_{\rm R}(\theta,T)$ shifts to the higher field side with increasing the angle $\theta$.
More precisely, the critical field of $\mu_0H_{\rm R}(0, T\!<\!1~{\rm K})=11.2$~T obtained here is slightly lower than the previously reported values~\cite{levy2005magnetic,levy2007acute,aoki2011ferromagnetic,hardy2011transverse}. 
The inset of Fig.~\ref{theta_zero_TCP_HR}(a) compares the critical field defined in the ascending (open squares) and descending (open circles) fields. 
A small but distinct hysteresis appears and grows in amplitude on cooling below 1.5~K.

Figures \ref{theta_zero_TCP_HR}(b) and \ref{theta_zero_TCP_HR}(c) show the temperature evolution of the amplitude of the peak in $d\tilde{M}/dH$  at $H_{\rm R}(\theta)$ and the transition width defined by the full width at the half maximum, respectively, measured at several $\theta$.
A remarkable weakening of the transition is evident above  2 K, the  temperature which is close to $T_{\rm TCP}$ reported previously~\cite{gourgout2016collapse}.
The transition width for $\theta\geq 0.79$$^\circ$ shows a significant broadening above 2~K.
At $\theta=0$, by contrast, the transition remains relatively sharp even near 6~K, suggesting that the first-order-like behavior persists up to this temperature.

Figure \ref{MH_deg_raw} shows the magnetization curves $M(H)$ of URhGe near $H_{\rm R}(\theta)$ for $\theta$ values from 0.79$^\circ$ to 5.64$^\circ$  measured at (a) 0.25, (b) 3, and  (c) 6 K, together with their differential curves $dM/dH$ for (d) 0.25, (e) 3, and  (f) 6 K.
Note that the torque component is properly subtracted for the results at $\theta > 0.1^\circ$.
Comparing Figs.~\ref{MH_deg_raw}(d)-\ref{MH_deg_raw}(f) with Figs.~\ref{Torque-theta-T_c}(d-f), one can see that $dM/dH$ shows qualitatively the same behavior with $-d(\Delta$$C_{\tau}/H)$$/dH$; the jump in $M_b$  is correlated to the negative jump in $M_c$. 
In particular, both data yield the same critical field $H_{\rm R}(\theta,T)$.

For comparison, $\tilde{M}(H)$ and $d\tilde{M}/dH$ for $\theta = 0$$^\circ$ are also plotted in these figures.
Whereas $\tilde{M}(H)$ agrees with $M(H)$ for $\theta\ge 0.79^\circ$ at fields above $H_{\rm R}$, 
an apparent disparity is evident below $H_{\rm R}$;
 $\tilde{M}(H)$ appears to be underestimated.
We attribute this problem to an incomplete subtraction of the torque component in the $\theta=0^\circ$ condition. 
A difficulty is that the vertical field gradient $G$ produces a small $c$-axis component of the magnetic field on the sample~\cite{note} and slightly deflects the field angle, accordingly. We estimate the angle shift to be $\sim -0.05^\circ$ at $G=8$~T/m. Even such a tiny change in $\theta$, however, causes a significant effect at $\theta\approx0$$^\circ$ because the torque component for $H<H_{\rm R}$ shows very strong $\theta$ variation there (Fig.~\ref{Capa-deg_0p6K_c}).
As a consequence, the condition of the torque component being independent of $G$ fails, resulting in an incomplete torque subtraction for $H<H_{\rm R}$ and an overestimate of the magnetization jump. 

In order to get a reliable estimation of the magnetization jump for $\theta=0$$^\circ$, we plot 
the metamagnetic jump $\Delta M$ for the field direction $\theta = 0.79$$^\circ$, 1.65$^\circ$ and 3.64$^\circ$ in the inset of Figs. \ref{MH_deg_raw}(a) and \ref{MH_deg_raw}(b). 
In both plots, $\Delta M$ shows a gradual angular variation, and its linear extrapolation to $\theta=0$$^\circ$ gives 
$\Delta M\sim 0.09$ $\mu_{\rm B}$/f.u. at 0.25 K and $\sim0.07$ $\mu_{\rm B}$/f.u. at 3 K.
One sees that $\Delta \tilde{M}$ for $\theta = 0$$^\circ$ (open circles) is about 1.5 times overestimated for both $T=0.25$  and 3 K. 
In what follows, accordingly, we reduce the peak value of $d\tilde{M}/dH$ by a factor 0.7 in the discussion of the angular variation of the transition. 
We note that thus corrected amplitude of the magnetization jump at 0.25~K for $\theta=0$$^\circ$ is in good agreement with the previous estimate by $T\rightarrow 0$ extrapolation of the $M_b(T)$ data~\cite{hardy2011transverse}.
\color{black}

\section{Discussion}

We employ these data of $dM/dH$ and $d(\Delta$$C_{\tau})$$/dH$ for construction of the URhGe wing structure phase diagram. As $H(\theta)$ passes through the first-order wing plane at a fixed $T$ in the $T-H_b-H_c$ phase diagram (see Fig.~\ref{3D_v2_c}(c)), $dM/dH$ as well as $|d(\Delta$$C_{\tau})$$/dH|$ exhibit a peak. Mapping those peak positions to the $T-H_b-H_c$ space then provides the wing phase diagram. 
Figure \ref{Wing-H_c_Mag} shows the contour plot of $dM/dH$ around the FM wing structure of  URhGe in the $H_b$-$H_c$ plane  at various temperatures.
Dotted lines indicate the traces of the field sweep at fixed angles $\theta$, along which the magnetization data were obtained. 
Green dots on the contour plots represent the peak position of $dM/dH$ measured at $\theta = 0$$^\circ$, 0.79$^\circ$, 1.65$^\circ$, 3.64$^\circ$, and 5.64$^\circ$, and solid lines are guides to the eye.
One can see that the bright arc in  Fig.~\ref{Wing-H_c_Mag}, i.e., the first-order transition region, becomes narrower as $T$ increases. Above 4.2~K, the bright spot can only be seen at $H_c=0$, indicating the first-order transition is confined to the narrow region. Similar plots can also be obtained from the $d(\Delta$$C_{\tau})$$/dH$ data and the results are shown in Fig.~\ref{Wing-H_c_torque}. 
Note that the data points are absent at $H_c=0$ in Fig.~\ref{3D_v2_c}(b) because the torque component vanishes there due to a ferromagnetic domain formation. 
These plots thus represent cuts of the wing plane at various temperatures (see Fig.~\ref{3D_v2_c}(c)), indicating that the wing planes are slightly warped. 

\begin{figure}[t]
\begin{center}
\includegraphics[width=0.95\linewidth]{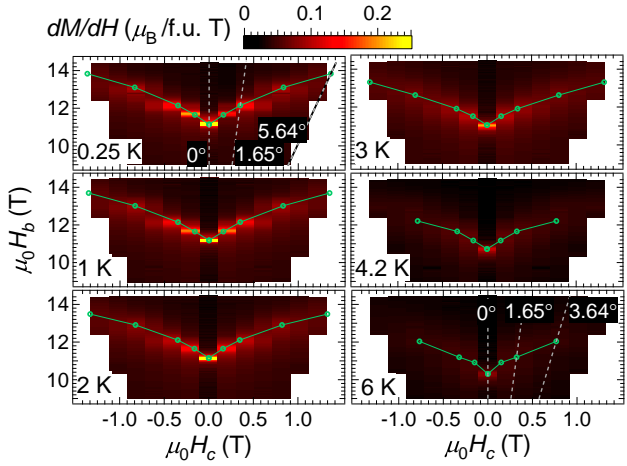}
\caption{ 
(color online). The contour plot of $dM/dH$ near the FM wing structure of  URhGe in the $H_b$-$H_c$ plane for various temperatures 0.25, 1, 2, 3, 4.2, and 6 K. Green dots represent the peak position of $dM/dH$ obtained at $\theta$ = 0$^\circ$, 0.79$^\circ$, 1.65$^\circ$, 3.64$^\circ$, and 5.64$^\circ$, and the green solid lines are guide to the eye. Mirrored copy data are plotted for $\theta<0$$^\circ$.
\label{Wing-H_c_Mag}}
\end{center}
\end{figure}
\begin{figure}[t]
\begin{center}
\includegraphics[width=0.95\linewidth]{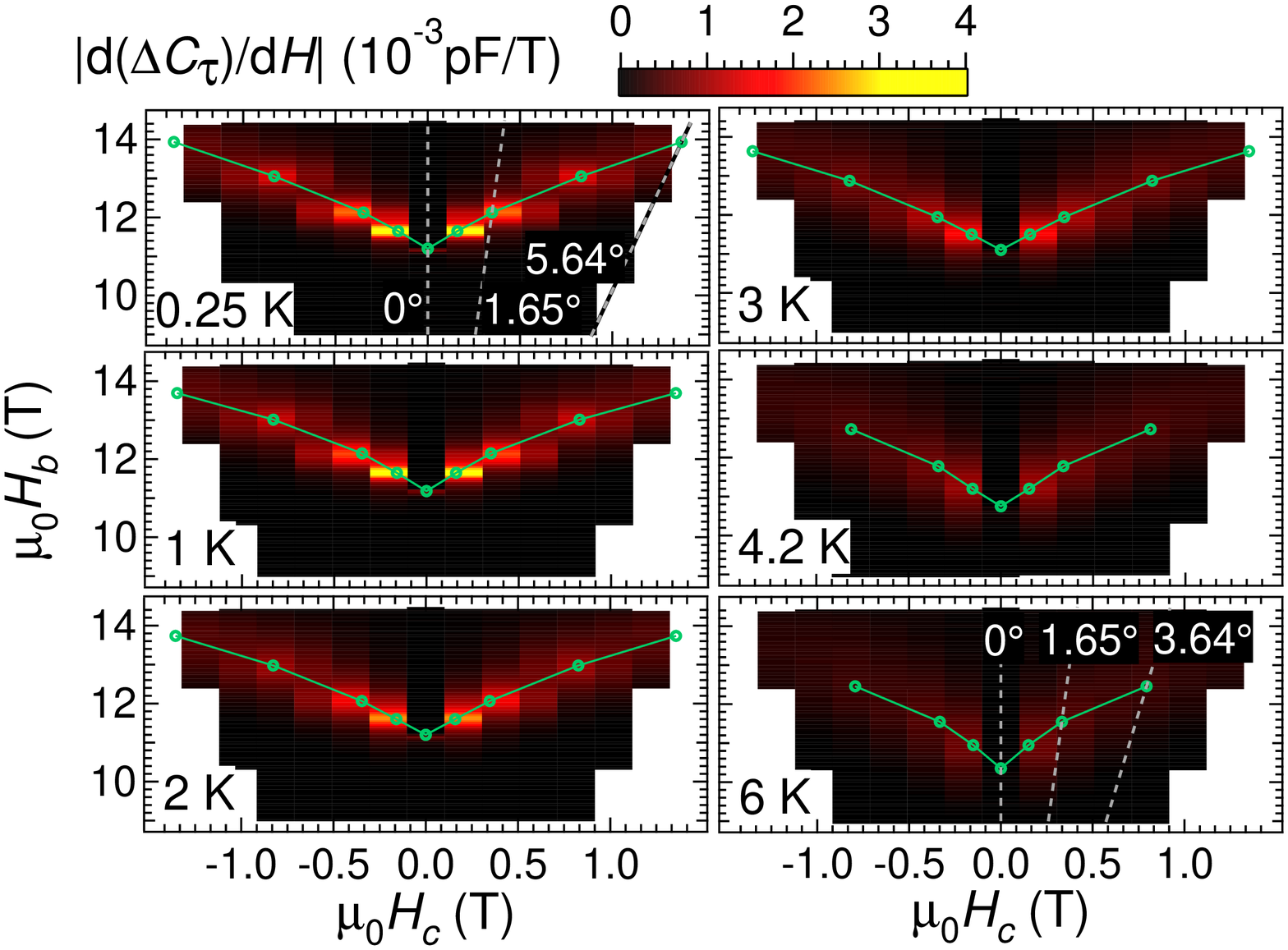}
\caption{ 
(color online). The contour plot of $|d(\Delta$$C_{\tau})$$/dH|$ near the FM wing structure of URhGe in the $H_b$-$H_c$ plane for various temperatures 0.25, 1, 2, 3, 4.2, and 6 K. Green dots represent the peak position of $|d(\Delta$$C_{\tau})$$/dH|$ obtained at $\theta$ = 0$^\circ$, 0.79$^\circ$, 1.65$^\circ$, 3.64$^\circ$, and 5.64$^\circ$, and the green solid lines are guide to the eye. Mirrored copy data are plotted for $\theta<0$$^\circ$.
\label{Wing-H_c_torque}}
\end{center}
\end{figure}
\begin{figure}[t]
\begin{center}
\includegraphics[width=0.95\linewidth]{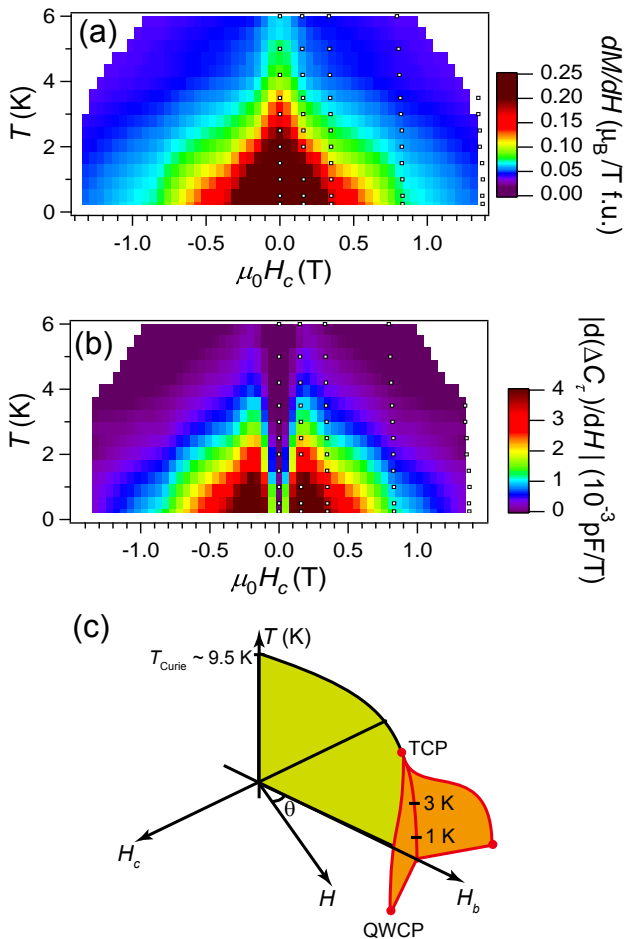}
\caption{ 
(color online). The color contour plot of the peak amplitude of (a) $dM/dH$ and (b) $|d(\Delta$$C_{\tau})$$/dH|$, projected on the $T-H_c$ plane. These plots  give imaging of the wing plane, viewed from the $H_b$ axis.   These plots are constructed from the data obtained at $T = 0.5$, 1.5, 2.5, 3.5, and 5 K (not shown), in addition to those given in Figs. \ref{Wing-H_c_Mag} and \ref{Wing-H_c_torque}.  The white dots in these figures show the data points from which the color mappings are generated.  A schematic $T-H_b-H_c$ phase diagram is given in (c).
\label{3D_v2_c}}
\end{center}
\end{figure}

Figure \ref{3D_v2_c} shows the color contour plot of the peak amplitude of (a) $dM/dH$ and (b) $|d(\Delta$$C_{\tau})$$/dH|$, projected on the $T-H_c$ plane. 
These plots  provide imaging of the wing plane viewed from the $H_b$ axis. 
Overall, the wing plane is bell shaped and steeply extends to higher temperatures above $\sim$4~K  at $H_c=0$. This feature of the wing plane is in agreement with the phenomenological analysis that three second-order transition lines meet at TCP tangentially~\cite{PhysRevB.94.060410}.

In a prototypical ferromagnet,  a first-order transition changes into a second-order one at the edge of the wing plane. One therefore expects that $|d(\Delta$$C_{\tau})$$/dH(\theta)|$, the field derivative of the OP, becomes divergent on the line connecting TCP and QWCP \cite{PhysRevB.86.024428}. 
Unlike the expectation, however, the peak amplitude of $|d(\Delta$$C_{\tau})$$/dH(\theta)|$ of URhGe decreases progressively as $\theta$ increases, making it somewhat difficult to define the wing edge from these data. This observation, along with the smallness of the hysteresis in $H_{\rm R}$, demonstrate the weak nature of the first-order transition in this compound.
Nevertheless, from Fig.~\ref{3D_v2_c}(b) we may judge that the wing plane extends to $\mu_0H_c\sim$1.1~T at $T\rightarrow 0$, because outside this range the landscape of $|d(\Delta$$C_{\tau})$$/dH(\theta)|$ becomes suddenly flat and low. Thus the location of  QWCP is estimated to be $\mu_0H_c\sim$1.1~T and $\mu_0H_b\sim$13.5~T.

Similar difficulty exists in the determination of TCP. Since $|d(\Delta$$C_{\tau})$$/dH(\theta)|$ has poor sensitivity to detect TCP at $\theta=0$ ($H_c=0$) because of the ferromagnetic domain issue, we inspect the $dM_b/dH$ data. One should keep in mind that $M_b$ is not the OP of the phase transition under consideration. We therefore need some theoretical inputs to discuss the phase transition by the $dM/dH$ data near $\theta=0$$^\circ$.
Up to now, no established microscopic theory is at hand for the field-induced phase transition in URhGe. We thus rely on the phenomenological model~\cite{PhysRevB.91.014506} that treats the phase transition of an Ising ferromagnet in a magnetic field perpendicular to the spontaneous magnetization. According to the theory, $M_b$ can be expressed in terms of $M_c$ as
\begin{equation}
M_b=\frac{H_b}{2(\alpha+\beta M_c^2)},
\end{equation}
where $\alpha$ and $\beta$ are the coefficients of the $M_b^2$ and the $M_c^2M_b^2$ terms in the Landau free energy expansion, respectively. 
A first-order spin reorientation transition is predicted by this model when $\beta$ exceeds a certain critical value~\cite{PhysRevB.91.014506}.
As described in the next paragraph, we can see from this equation how $dM_b/dH_b$ at the transition evolves with $T$ in the $T-H_b$ plane; $dM_b/dH_b$ diverges at the transition for $T\leq T_{\rm TCP}$, whereas it does not for $T> T_{\rm TCP}$.

At a second-order transition point above TCP, $M_c$ on the $H_c=0$ plane develops as  $M_c\propto \sqrt{T_{\rm C}(H_b)-T}$, where $T_{\rm C}(H_b)$ is given by
\begin{equation}
T_{\rm C}(H_b)=T_{\rm C}(0)-A\beta H_b^2,
\end{equation}
with $A$ being a constant~\cite{PhysRevB.91.014506}. Even though $M_c$ shows an infinite change of slope at  $T_{\rm C}(H_b)$, $dM_b/dH_b$ does not exhibit a strong singularity; 
from Eqs. (1) and (2), $M_b$ would only exhibit a finite change of slope as a function of $T$ or $H_b$. 
This feature of $M_b(T)$ can indeed be seen in the magnetization data measured in various fields $H_b$~\cite{hardy2011transverse}.
By contrast, just at $T=T_{\rm TCP}$, $M_c \propto  (T_{\rm C}(H_b)-T)^{1/4}$ because the $M_c^4$ term in the renormalized free energy vanishes. In this case, $dM_b/dH_b$ would diverge as $(T_{\rm C}(H_b)-T)^{-1/2}$ because a square-root singularity remains in $M_b$.
Below TCP, $dM_b/dH_b$ diverges as well, reflecting a finite jump of $M_c$ at the first-order transition.

By looking at the $dM/dH$ contour plot in Fig.~\ref{3D_v2_c}(a), we find that
the peak amplitude for $\theta=0$$^\circ$ becomes progressively smaller with increasing $T$ above 2~K.  It can be seen, however, that $dM_b/dH_b$ in Fig.~\ref{MH_deg_raw}(f) still exhibits a rather sharp peak, i.e. divergent behavior, at 6~K. This fact suggests that the first-order nature of the transition persists up to this temperature.
We note that this important feature of the transition is observable only in a very narrow angular window of $|\theta|<0.8^\circ$.


Up to now, there have been a few reports regarding the location of TCP in URhGe.
In the $^{73}$Ge NMR spectra study performed in a field of 12~T applied parallel to the $b$ axis, a phase separation of the FM and the paramagnetic states, the fingerprint of the first-order transition,  can be seen at least up to 4.3~K, giving rather strong evidence that $T_{\rm TCP}$ is well above this temperature~\cite{kotegawa201573ge}. 
By contrast, the thermoelectric power experiment claims much lower TCP temperature of 2~K~\cite{gourgout2016collapse}.
The wing structure phase diagram (Fig.~\ref{3D_v2_c}) obtained in the present experiment is consistent with the NMR results. It should be noticed that a misalignment of the magnetic field by $\sim 1^\circ$ from the $b$ axis would yield an incorrect estimate  of $T_{\rm TCP}\lesssim 3$~K.

Finally, some remarks are made regarding RSC in URhGe. The RSC in this system emerges not only near the quantum wing  critical point, but also along the first-order quantum phase transition line of the wing structure at $T=0$. Indeed, the zero-resistivity state of RSC at 50~mK occurs along the first-order transition line in the $H_b-H_c$ plane, terminating at QWCP~\cite{levy2005magnetic}.
A possible origin of this unusual phenomena has been attributed to longitudinal ($\parallel b$) magnetic fluctuations, and discussed in relation to a quantum TCP that can be expected when $T_{\rm TCP}$ is very low~\cite{levy2005magnetic,levy2009coexistence,tokunaga2015reentrant}.
The present results reveal, however, that there is a large disparity between $T_{\rm TCP}>4$~K and $T_{\rm RSC}\approx 0.42$~K; $T_{\rm TCP}/T_{\rm RSC}\gtrsim 10$, indicating that the system is not close to a quantum TCP. In this regard, we point out that the first-order transition in this system is very weak in nature, as evidenced by a smallness in the hysteresis of the critical field as well as a rapid broadening of the transition with $T$. Such a weakness of the first-order transition might host substantial fluctuations even at low temperatures $T\ll T_{\rm TCP}$.

\section{Conclusion}
 
We have investigated the quantum phase transition of an Ising ferromagnet URhGe by means of 
 high-precision angle-resolved dc magnetization measurements in magnetic fields applied near the $b$ axis.
A first-order spin reorientation transition has been observed at low temperatures, accompanied by a small hysteresis in the critical field.
The temperature and angular variations of the transition observed in the magnetization as well as in the magnetic torque allow us to construct the three-dimensional  $T-H_c-H_b$ phase diagram, where $H_c$ ($\parallel c$) is the conjugate field parallel to the order parameter and $H_b$ is the $b$-axis component of the field that tunes $T_{\rm C}$ down to zero.
The tricritical point $T_{\rm TCP}$ is estimated to be located above 4~K in the $H_c=0$ plane.
On cooling below $T_{\rm TCP}$, a wing structure develops by increasing $|H_c|$. We have succeeded in directly determining the detailed profiles of the wing structure. The quantum wing critical points exist at $H_c=\pm 1.1$~T and $H_b=13.5$~T. Three second-order transition lines meet at $T_{\rm TCP}$ tangentially, so that a precise tuning of $H$ along the $b$ axis within 0.8$^\circ$ is needed to correctly determine the position of TCP.
The reentrant superconductivity in this system is not due to a quantum TCP~\cite{levy2007acute}, but is rather related to unusually weak nature of the first-order transition represented by a smallness of the hysteresis and a broadness of the transition.

\begin{acknowledgments}
The present work was supported in part by a Grant-in-Aid for Scientific Research on Innovative Areas ``J-Physics'' (15H05883)  and KAKENHI (15H03682) from MEXT.

\end{acknowledgments}


\bibliography{apssamp.bib}

\begin{thebibliography}{38}
\expandafter\ifx\csname natexlab\endcsname\relax\def\natexlab#1{#1}\fi
\expandafter\ifx\csname bibnamefont\endcsname\relax
  \def\bibnamefont#1{#1}\fi
\expandafter\ifx\csname bibfnamefont\endcsname\relax
  \def\bibfnamefont#1{#1}\fi
\expandafter\ifx\csname citenamefont\endcsname\relax
  \def\citenamefont#1{#1}\fi
\expandafter\ifx\csname url\endcsname\relax
  \def\url#1{\texttt{#1}}\fi
\expandafter\ifx\csname urlprefix\endcsname\relax\def\urlprefix{URL }\fi
\providecommand{\bibinfo}[2]{#2}
\providecommand{\eprint}[2][]{\url{#2}}

\bibitem[{\citenamefont{Saxena et~al.}(2000)\citenamefont{Saxena, Agarwal,
  Ahilan, Grosche, Haselwimmer, Steiner, Pugh, Walker, Julian, Monthoux
  et~al.}}]{saxena2000superconductivity}
\bibinfo{author}{\bibfnamefont{S.}~\bibnamefont{Saxena}},
  \bibinfo{author}{\bibfnamefont{P.}~\bibnamefont{Agarwal}},
  \bibinfo{author}{\bibfnamefont{K.}~\bibnamefont{Ahilan}},
  \bibinfo{author}{\bibfnamefont{F.}~\bibnamefont{Grosche}},
  \bibinfo{author}{\bibfnamefont{R.}~\bibnamefont{Haselwimmer}},
  \bibinfo{author}{\bibfnamefont{M.}~\bibnamefont{Steiner}},
  \bibinfo{author}{\bibfnamefont{E.}~\bibnamefont{Pugh}},
  \bibinfo{author}{\bibfnamefont{I.}~\bibnamefont{Walker}},
  \bibinfo{author}{\bibfnamefont{S.}~\bibnamefont{Julian}},
  \bibinfo{author}{\bibfnamefont{P.}~\bibnamefont{Monthoux}},
  \bibnamefont{et~al.}, \bibinfo{journal}{Nature}
  \textbf{\bibinfo{volume}{406}}, \bibinfo{pages}{587} (\bibinfo{year}{2000}).

\bibitem[{\citenamefont{Aoki et~al.}(2001)\citenamefont{Aoki, Huxley,
  Ressouche, Braithwaite, Flouquet, Brison, Lhotel, and
  Paulsen}}]{aoki2001coexistence}
\bibinfo{author}{\bibfnamefont{D.}~\bibnamefont{Aoki}},
  \bibinfo{author}{\bibfnamefont{A.}~\bibnamefont{Huxley}},
  \bibinfo{author}{\bibfnamefont{E.}~\bibnamefont{Ressouche}},
  \bibinfo{author}{\bibfnamefont{D.}~\bibnamefont{Braithwaite}},
  \bibinfo{author}{\bibfnamefont{J.}~\bibnamefont{Flouquet}},
  \bibinfo{author}{\bibfnamefont{J.-P.} \bibnamefont{Brison}},
  \bibinfo{author}{\bibfnamefont{E.}~\bibnamefont{Lhotel}}, \bibnamefont{and}
  \bibinfo{author}{\bibfnamefont{C.}~\bibnamefont{Paulsen}},
  \bibinfo{journal}{Nature} \textbf{\bibinfo{volume}{413}},
  \bibinfo{pages}{613} (\bibinfo{year}{2001}).

\bibitem[{\citenamefont{Huy et~al.}(2007)\citenamefont{Huy, Gasparini, De~Nijs,
  Huang, Klaasse, Gortenmulder, de~Visser, Hamann, G{\"o}rlach, and
  L{\"o}hneysen}}]{huy2007superconductivity}
\bibinfo{author}{\bibfnamefont{N.}~\bibnamefont{Huy}},
  \bibinfo{author}{\bibfnamefont{A.}~\bibnamefont{Gasparini}},
  \bibinfo{author}{\bibfnamefont{D.}~\bibnamefont{De~Nijs}},
  \bibinfo{author}{\bibfnamefont{Y.}~\bibnamefont{Huang}},
  \bibinfo{author}{\bibfnamefont{J.}~\bibnamefont{Klaasse}},
  \bibinfo{author}{\bibfnamefont{T.}~\bibnamefont{Gortenmulder}},
  \bibinfo{author}{\bibfnamefont{A.}~\bibnamefont{de~Visser}},
  \bibinfo{author}{\bibfnamefont{A.}~\bibnamefont{Hamann}},
  \bibinfo{author}{\bibfnamefont{T.}~\bibnamefont{G{\"o}rlach}},
  \bibnamefont{and} \bibinfo{author}{\bibfnamefont{H.~v.}
  \bibnamefont{L{\"o}hneysen}}, \bibinfo{journal}{Phys. Rev. Lett.}
  \textbf{\bibinfo{volume}{99}}, \bibinfo{pages}{067006}
  (\bibinfo{year}{2007}).

\bibitem[{\citenamefont{Hattori et~al.}(2012)\citenamefont{Hattori, Ihara,
  Nakai, Ishida, Tada, Fujimoto, Kawakami, Osaki, Deguchi, Sato
  et~al.}}]{hattori2012superconductivity}
\bibinfo{author}{\bibfnamefont{T.}~\bibnamefont{Hattori}},
  \bibinfo{author}{\bibfnamefont{Y.}~\bibnamefont{Ihara}},
  \bibinfo{author}{\bibfnamefont{Y.}~\bibnamefont{Nakai}},
  \bibinfo{author}{\bibfnamefont{K.}~\bibnamefont{Ishida}},
  \bibinfo{author}{\bibfnamefont{Y.}~\bibnamefont{Tada}},
  \bibinfo{author}{\bibfnamefont{S.}~\bibnamefont{Fujimoto}},
  \bibinfo{author}{\bibfnamefont{N.}~\bibnamefont{Kawakami}},
  \bibinfo{author}{\bibfnamefont{E.}~\bibnamefont{Osaki}},
  \bibinfo{author}{\bibfnamefont{K.}~\bibnamefont{Deguchi}},
  \bibinfo{author}{\bibfnamefont{N.}~\bibnamefont{Sato}}, \bibnamefont{et~al.},
  \bibinfo{journal}{Phys. Rev. Lett.} \textbf{\bibinfo{volume}{108}},
  \bibinfo{pages}{066403} (\bibinfo{year}{2012}).

\bibitem[{\citenamefont{Ohta et~al.}(2010)\citenamefont{Ohta, Hattori, Ishida,
  Nakai, Osaki, Deguchi, K.~Sato, and Satoh}}]{ohta2010microscopic}
\bibinfo{author}{\bibfnamefont{T.}~\bibnamefont{Ohta}},
  \bibinfo{author}{\bibfnamefont{T.}~\bibnamefont{Hattori}},
  \bibinfo{author}{\bibfnamefont{K.}~\bibnamefont{Ishida}},
  \bibinfo{author}{\bibfnamefont{Y.}~\bibnamefont{Nakai}},
  \bibinfo{author}{\bibfnamefont{E.}~\bibnamefont{Osaki}},
  \bibinfo{author}{\bibfnamefont{K.}~\bibnamefont{Deguchi}},
  \bibinfo{author}{\bibfnamefont{N.}~\bibnamefont{K.~Sato}}, \bibnamefont{and}
  \bibinfo{author}{\bibfnamefont{I.}~\bibnamefont{Satoh}}, \bibinfo{journal}{J.
  Phys. Soc. Jpn.} \textbf{\bibinfo{volume}{79}}, \bibinfo{pages}{023707}
  (\bibinfo{year}{2010}).

\bibitem[{\citenamefont{de~Visser et~al.}(2009)\citenamefont{de~Visser, Huy,
  Gasparini, de~Nijs, Andreica, Baines, and Amato}}]{PhysRevLett.102.167003}
\bibinfo{author}{\bibfnamefont{A.}~\bibnamefont{de~Visser}},
  \bibinfo{author}{\bibfnamefont{N.~T.} \bibnamefont{Huy}},
  \bibinfo{author}{\bibfnamefont{A.}~\bibnamefont{Gasparini}},
  \bibinfo{author}{\bibfnamefont{D.~E.} \bibnamefont{de~Nijs}},
  \bibinfo{author}{\bibfnamefont{D.}~\bibnamefont{Andreica}},
  \bibinfo{author}{\bibfnamefont{C.}~\bibnamefont{Baines}}, \bibnamefont{and}
  \bibinfo{author}{\bibfnamefont{A.}~\bibnamefont{Amato}},
  \bibinfo{journal}{Phys. Rev. Lett.} \textbf{\bibinfo{volume}{102}},
  \bibinfo{pages}{167003} (\bibinfo{year}{2009}).

\bibitem[{\citenamefont{Kotegawa et~al.}(2005)\citenamefont{Kotegawa, Harada,
  Kawasaki, Kawasaki, Kitaoka, Haga, Yamamoto, Onuki, Itoh, Haller
  et~al.}}]{kotegawaJPSJ.74.705coexist}
\bibinfo{author}{\bibfnamefont{H.}~\bibnamefont{Kotegawa}},
  \bibinfo{author}{\bibfnamefont{A.}~\bibnamefont{Harada}},
  \bibinfo{author}{\bibfnamefont{S.}~\bibnamefont{Kawasaki}},
  \bibinfo{author}{\bibfnamefont{Y.}~\bibnamefont{Kawasaki}},
  \bibinfo{author}{\bibfnamefont{Y.}~\bibnamefont{Kitaoka}},
  \bibinfo{author}{\bibfnamefont{Y.}~\bibnamefont{Haga}},
  \bibinfo{author}{\bibfnamefont{E.}~\bibnamefont{Yamamoto}},
  \bibinfo{author}{\bibfnamefont{Y.}~\bibnamefont{Onuki}},
  \bibinfo{author}{\bibfnamefont{K.~M.} \bibnamefont{Itoh}},
  \bibinfo{author}{\bibfnamefont{E.}~\bibnamefont{Haller}},
  \bibnamefont{et~al.}, \bibinfo{journal}{J. Phys. Soc. Jpn.}
  \textbf{\bibinfo{volume}{74}}, \bibinfo{pages}{705} (\bibinfo{year}{2005}).

\bibitem[{\citenamefont{Aoki et~al.}(2009)\citenamefont{Aoki, Matsuda, Taufour,
  Hassinger, Knebel, and Flouquet}}]{200915989}
\bibinfo{author}{\bibfnamefont{D.}~\bibnamefont{Aoki}},
  \bibinfo{author}{\bibfnamefont{T.~D.} \bibnamefont{Matsuda}},
  \bibinfo{author}{\bibfnamefont{V.}~\bibnamefont{Taufour}},
  \bibinfo{author}{\bibfnamefont{E.}~\bibnamefont{Hassinger}},
  \bibinfo{author}{\bibfnamefont{G.}~\bibnamefont{Knebel}}, \bibnamefont{and}
  \bibinfo{author}{\bibfnamefont{J.}~\bibnamefont{Flouquet}},
  \bibinfo{journal}{J. Phys. Soc. Jpn.} \textbf{\bibinfo{volume}{78}},
  \bibinfo{pages}{113709} (\bibinfo{year}{2009}).

\bibitem[{\citenamefont{L{\'e}vy et~al.}(2009)\citenamefont{L{\'e}vy, Sheikin,
  Grenier, Marcenat, and Huxley}}]{levy2009coexistence}
\bibinfo{author}{\bibfnamefont{F.}~\bibnamefont{L{\'e}vy}},
  \bibinfo{author}{\bibfnamefont{I.}~\bibnamefont{Sheikin}},
  \bibinfo{author}{\bibfnamefont{B.}~\bibnamefont{Grenier}},
  \bibinfo{author}{\bibfnamefont{C.}~\bibnamefont{Marcenat}}, \bibnamefont{and}
  \bibinfo{author}{\bibfnamefont{A.}~\bibnamefont{Huxley}},
  \bibinfo{journal}{J. Phys.: Condens. Matter} \textbf{\bibinfo{volume}{21}},
  \bibinfo{pages}{164211} (\bibinfo{year}{2009}).

\bibitem[{\citenamefont{Tokunaga et~al.}(2015)\citenamefont{Tokunaga, Aoki,
  Mayaffre, Kr{\"a}mer, Julien, Berthier, Horvati{\'c}, Sakai, Kambe, and
  Araki}}]{tokunaga2015reentrant}
\bibinfo{author}{\bibfnamefont{Y.}~\bibnamefont{Tokunaga}},
  \bibinfo{author}{\bibfnamefont{D.}~\bibnamefont{Aoki}},
  \bibinfo{author}{\bibfnamefont{H.}~\bibnamefont{Mayaffre}},
  \bibinfo{author}{\bibfnamefont{S.}~\bibnamefont{Kr{\"a}mer}},
  \bibinfo{author}{\bibfnamefont{M.-H.} \bibnamefont{Julien}},
  \bibinfo{author}{\bibfnamefont{C.}~\bibnamefont{Berthier}},
  \bibinfo{author}{\bibfnamefont{M.}~\bibnamefont{Horvati{\'c}}},
  \bibinfo{author}{\bibfnamefont{H.}~\bibnamefont{Sakai}},
  \bibinfo{author}{\bibfnamefont{S.}~\bibnamefont{Kambe}}, \bibnamefont{and}
  \bibinfo{author}{\bibfnamefont{S.}~\bibnamefont{Araki}},
  \bibinfo{journal}{Phys. Rev. Lett.} \textbf{\bibinfo{volume}{114}},
  \bibinfo{pages}{216401} (\bibinfo{year}{2015}).

\bibitem[{\citenamefont{Taufour et~al.}(2011)\citenamefont{Taufour, Villaume,
  Aoki, Knebel, and Flouquet}}]{taufour1742-6596-273-1-012017}
\bibinfo{author}{\bibfnamefont{V.}~\bibnamefont{Taufour}},
  \bibinfo{author}{\bibfnamefont{A.}~\bibnamefont{Villaume}},
  \bibinfo{author}{\bibfnamefont{D.}~\bibnamefont{Aoki}},
  \bibinfo{author}{\bibfnamefont{G.}~\bibnamefont{Knebel}}, \bibnamefont{and}
  \bibinfo{author}{\bibfnamefont{J.}~\bibnamefont{Flouquet}},
  \bibinfo{journal}{J. Phys. Conf. Ser.} \textbf{\bibinfo{volume}{273}},
  \bibinfo{pages}{012017} (\bibinfo{year}{2011}).

\bibitem[{\citenamefont{Tran et~al.}(1998)\citenamefont{Tran, Tro{\'{c}}, and
  Andr{\'{e}}}}]{TRAN199881}
\bibinfo{author}{\bibfnamefont{V.}~\bibnamefont{Tran}},
  \bibinfo{author}{\bibfnamefont{R.}~\bibnamefont{Tro{\'{c}}}},
  \bibnamefont{and}
  \bibinfo{author}{\bibfnamefont{G.}~\bibnamefont{Andr{\'{e}}}},
  \bibinfo{journal}{J. Magn. Magn. Mater.} \textbf{\bibinfo{volume}{186}},
  \bibinfo{pages}{81 } (\bibinfo{year}{1998}).

\bibitem[{\citenamefont{L{\'e}vy et~al.}(2005)\citenamefont{L{\'e}vy, Sheikin,
  Grenier, and Huxley}}]{levy2005magnetic}
\bibinfo{author}{\bibfnamefont{F.}~\bibnamefont{L{\'e}vy}},
  \bibinfo{author}{\bibfnamefont{I.}~\bibnamefont{Sheikin}},
  \bibinfo{author}{\bibfnamefont{B.}~\bibnamefont{Grenier}}, \bibnamefont{and}
  \bibinfo{author}{\bibfnamefont{A.~D.} \bibnamefont{Huxley}},
  \bibinfo{journal}{Science} \textbf{\bibinfo{volume}{309}},
  \bibinfo{pages}{1343} (\bibinfo{year}{2005}).

\bibitem[{\citenamefont{Hardy et~al.}(2011)\citenamefont{Hardy, Aoki, Meingast,
  Schweiss, Burger, L{\"o}hneysen, and Flouquet}}]{hardy2011transverse}
\bibinfo{author}{\bibfnamefont{F.}~\bibnamefont{Hardy}},
  \bibinfo{author}{\bibfnamefont{D.}~\bibnamefont{Aoki}},
  \bibinfo{author}{\bibfnamefont{C.}~\bibnamefont{Meingast}},
  \bibinfo{author}{\bibfnamefont{P.}~\bibnamefont{Schweiss}},
  \bibinfo{author}{\bibfnamefont{P.}~\bibnamefont{Burger}},
  \bibinfo{author}{\bibfnamefont{H.~v.} \bibnamefont{L{\"o}hneysen}},
  \bibnamefont{and} \bibinfo{author}{\bibfnamefont{J.}~\bibnamefont{Flouquet}},
  \bibinfo{journal}{Phys. Rev. B} \textbf{\bibinfo{volume}{83}},
  \bibinfo{pages}{195107} (\bibinfo{year}{2011}).

\bibitem[{\citenamefont{Pfeuty}(1970)}]{PFEUTY197079}
\bibinfo{author}{\bibfnamefont{P.}~\bibnamefont{Pfeuty}},
  \bibinfo{journal}{Ann. Phys.} \textbf{\bibinfo{volume}{57}},
  \bibinfo{pages}{79 } (\bibinfo{year}{1970}).

\bibitem[{\citenamefont{Aoki and Flouquet}(2014)}]{aoki2014superconductivity}
\bibinfo{author}{\bibfnamefont{D.}~\bibnamefont{Aoki}} \bibnamefont{and}
  \bibinfo{author}{\bibfnamefont{J.}~\bibnamefont{Flouquet}},
  \bibinfo{journal}{J. Phys. Soc. Jpn.} \textbf{\bibinfo{volume}{83}},
  \bibinfo{pages}{061011} (\bibinfo{year}{2014}).

\bibitem[{\citenamefont{Brando et~al.}(2016)\citenamefont{Brando, Belitz,
  Grosche, and Kirkpatrick}}]{RevModPhys.88.025006}
\bibinfo{author}{\bibfnamefont{M.}~\bibnamefont{Brando}},
  \bibinfo{author}{\bibfnamefont{D.}~\bibnamefont{Belitz}},
  \bibinfo{author}{\bibfnamefont{F.~M.} \bibnamefont{Grosche}},
  \bibnamefont{and} \bibinfo{author}{\bibfnamefont{T.~R.}
  \bibnamefont{Kirkpatrick}}, \bibinfo{journal}{Rev. Mod. Phys.}
  \textbf{\bibinfo{volume}{88}}, \bibinfo{pages}{025006}
  (\bibinfo{year}{2016}).

\bibitem[{\citenamefont{Belitz et~al.}(1999)\citenamefont{Belitz, Kirkpatrick,
  and Vojta}}]{belitz1999first}
\bibinfo{author}{\bibfnamefont{D.}~\bibnamefont{Belitz}},
  \bibinfo{author}{\bibfnamefont{T.~R.} \bibnamefont{Kirkpatrick}},
  \bibnamefont{and} \bibinfo{author}{\bibfnamefont{T.}~\bibnamefont{Vojta}},
  \bibinfo{journal}{Phys. Rev. Lett.} \textbf{\bibinfo{volume}{82}},
  \bibinfo{pages}{4707} (\bibinfo{year}{1999}).

\bibitem[{\citenamefont{Belitz et~al.}(2005)\citenamefont{Belitz, Kirkpatrick,
  and Rollb\"uhler}}]{PhysRevLett.94.247205}
\bibinfo{author}{\bibfnamefont{D.}~\bibnamefont{Belitz}},
  \bibinfo{author}{\bibfnamefont{T.~R.} \bibnamefont{Kirkpatrick}},
  \bibnamefont{and}
  \bibinfo{author}{\bibfnamefont{J.}~\bibnamefont{Rollb\"uhler}},
  \bibinfo{journal}{Phys. Rev. Lett.} \textbf{\bibinfo{volume}{94}},
  \bibinfo{pages}{247205} (\bibinfo{year}{2005}).

\bibitem[{\citenamefont{Kirkpatrick and Belitz}(2015)}]{PhysRevLett.115.020402}
\bibinfo{author}{\bibfnamefont{T.~R.} \bibnamefont{Kirkpatrick}}
  \bibnamefont{and} \bibinfo{author}{\bibfnamefont{D.}~\bibnamefont{Belitz}},
  \bibinfo{journal}{Phys. Rev. Lett.} \textbf{\bibinfo{volume}{115}},
  \bibinfo{pages}{020402} (\bibinfo{year}{2015}).

\bibitem[{\citenamefont{Taufour et~al.}(2010)\citenamefont{Taufour, Aoki,
  Knebel, and Flouquet}}]{taufour2010tricritical}
\bibinfo{author}{\bibfnamefont{V.}~\bibnamefont{Taufour}},
  \bibinfo{author}{\bibfnamefont{D.}~\bibnamefont{Aoki}},
  \bibinfo{author}{\bibfnamefont{G.}~\bibnamefont{Knebel}}, \bibnamefont{and}
  \bibinfo{author}{\bibfnamefont{J.}~\bibnamefont{Flouquet}},
  \bibinfo{journal}{Phys. Rev. Lett.} \textbf{\bibinfo{volume}{105}},
  \bibinfo{pages}{217201} (\bibinfo{year}{2010}).

\bibitem[{\citenamefont{Pfleiderer and Huxley}(2002)}]{pfleiderer2002pressure}
\bibinfo{author}{\bibfnamefont{C.}~\bibnamefont{Pfleiderer}} \bibnamefont{and}
  \bibinfo{author}{\bibfnamefont{A.~D.} \bibnamefont{Huxley}},
  \bibinfo{journal}{Phys. Rev. Lett.} \textbf{\bibinfo{volume}{89}},
  \bibinfo{pages}{147005} (\bibinfo{year}{2002}).

\bibitem[{\citenamefont{Kotegawa et~al.}(2011)\citenamefont{Kotegawa, Taufour,
  Aoki, Knebel, and Flouquet}}]{kotegawa2011evolution}
\bibinfo{author}{\bibfnamefont{H.}~\bibnamefont{Kotegawa}},
  \bibinfo{author}{\bibfnamefont{V.}~\bibnamefont{Taufour}},
  \bibinfo{author}{\bibfnamefont{D.}~\bibnamefont{Aoki}},
  \bibinfo{author}{\bibfnamefont{G.}~\bibnamefont{Knebel}}, \bibnamefont{and}
  \bibinfo{author}{\bibfnamefont{J.}~\bibnamefont{Flouquet}},
  \bibinfo{journal}{J. Phys. Soc. Jpn.} \textbf{\bibinfo{volume}{80}},
  \bibinfo{pages}{083703} (\bibinfo{year}{2011}).

\bibitem[{\citenamefont{Uhlarz et~al.}(2004)\citenamefont{Uhlarz, Pfleiderer,
  and Hayden}}]{uhlarz2004quantum}
\bibinfo{author}{\bibfnamefont{M.}~\bibnamefont{Uhlarz}},
  \bibinfo{author}{\bibfnamefont{C.}~\bibnamefont{Pfleiderer}},
  \bibnamefont{and} \bibinfo{author}{\bibfnamefont{S.~M.}
  \bibnamefont{Hayden}}, \bibinfo{journal}{Phys. Rev. Lett.}
  \textbf{\bibinfo{volume}{93}}, \bibinfo{pages}{256404}
  (\bibinfo{year}{2004}).

\bibitem[{\citenamefont{Shimizu et~al.}(2015)\citenamefont{Shimizu,
  Braithwaite, Salce, Combier, Aoki, Hering, Ramos, and
  Flouquet}}]{PhysRevB.91.125115}
\bibinfo{author}{\bibfnamefont{Y.}~\bibnamefont{Shimizu}},
  \bibinfo{author}{\bibfnamefont{D.}~\bibnamefont{Braithwaite}},
  \bibinfo{author}{\bibfnamefont{B.}~\bibnamefont{Salce}},
  \bibinfo{author}{\bibfnamefont{T.}~\bibnamefont{Combier}},
  \bibinfo{author}{\bibfnamefont{D.}~\bibnamefont{Aoki}},
  \bibinfo{author}{\bibfnamefont{E.~N.} \bibnamefont{Hering}},
  \bibinfo{author}{\bibfnamefont{S.~M.} \bibnamefont{Ramos}}, \bibnamefont{and}
  \bibinfo{author}{\bibfnamefont{J.}~\bibnamefont{Flouquet}},
  \bibinfo{journal}{Phys. Rev. B} \textbf{\bibinfo{volume}{91}},
  \bibinfo{pages}{125115} (\bibinfo{year}{2015}).

\bibitem[{\citenamefont{M{\'\i}{\v{s}}ek
  et~al.}(2017)\citenamefont{M{\'\i}{\v{s}}ek, Prokle{\v{s}}ka, Opletal,
  Proschek, Ka{\v{s}}til, Kamar{\'a}d, and Sechovsk{\'y}}}]{mivsek2017pressure}
\bibinfo{author}{\bibfnamefont{M.}~\bibnamefont{M{\'\i}{\v{s}}ek}},
  \bibinfo{author}{\bibfnamefont{J.}~\bibnamefont{Prokle{\v{s}}ka}},
  \bibinfo{author}{\bibfnamefont{P.}~\bibnamefont{Opletal}},
  \bibinfo{author}{\bibfnamefont{P.}~\bibnamefont{Proschek}},
  \bibinfo{author}{\bibfnamefont{J.}~\bibnamefont{Ka{\v{s}}til}},
  \bibinfo{author}{\bibfnamefont{J.}~\bibnamefont{Kamar{\'a}d}},
  \bibnamefont{and}
  \bibinfo{author}{\bibfnamefont{V.}~\bibnamefont{Sechovsk{\'y}}},
  \bibinfo{journal}{AIP Advances} \textbf{\bibinfo{volume}{7}},
  \bibinfo{pages}{055712} (\bibinfo{year}{2017}).

\bibitem[{\citenamefont{Aoki et~al.}(2011)\citenamefont{Aoki, Combier, Taufour,
  D.~Matsuda, Knebel, Kotegawa, and Flouquet}}]{aoki2011ferromagnetic}
\bibinfo{author}{\bibfnamefont{D.}~\bibnamefont{Aoki}},
  \bibinfo{author}{\bibfnamefont{T.}~\bibnamefont{Combier}},
  \bibinfo{author}{\bibfnamefont{V.}~\bibnamefont{Taufour}},
  \bibinfo{author}{\bibfnamefont{T.}~\bibnamefont{D.~Matsuda}},
  \bibinfo{author}{\bibfnamefont{G.}~\bibnamefont{Knebel}},
  \bibinfo{author}{\bibfnamefont{H.}~\bibnamefont{Kotegawa}}, \bibnamefont{and}
  \bibinfo{author}{\bibfnamefont{J.}~\bibnamefont{Flouquet}},
  \bibinfo{journal}{J. Phys. Soc. Jpn.} \textbf{\bibinfo{volume}{80}},
  \bibinfo{pages}{094711} (\bibinfo{year}{2011}).

\bibitem[{\citenamefont{L{\'e}vy et~al.}(2007)\citenamefont{L{\'e}vy, Sheikin,
  and Huxley}}]{levy2007acute}
\bibinfo{author}{\bibfnamefont{F.}~\bibnamefont{L{\'e}vy}},
  \bibinfo{author}{\bibfnamefont{I.}~\bibnamefont{Sheikin}}, \bibnamefont{and}
  \bibinfo{author}{\bibfnamefont{A.}~\bibnamefont{Huxley}},
  \bibinfo{journal}{Nat. Phys.} \textbf{\bibinfo{volume}{3}},
  \bibinfo{pages}{460} (\bibinfo{year}{2007}).

\bibitem[{\citenamefont{Gourgout et~al.}(2016)\citenamefont{Gourgout, Pourret,
  Knebel, Aoki, Seyfarth, and Flouquet}}]{gourgout2016collapse}
\bibinfo{author}{\bibfnamefont{A.}~\bibnamefont{Gourgout}},
  \bibinfo{author}{\bibfnamefont{A.}~\bibnamefont{Pourret}},
  \bibinfo{author}{\bibfnamefont{G.}~\bibnamefont{Knebel}},
  \bibinfo{author}{\bibfnamefont{D.}~\bibnamefont{Aoki}},
  \bibinfo{author}{\bibfnamefont{G.}~\bibnamefont{Seyfarth}}, \bibnamefont{and}
  \bibinfo{author}{\bibfnamefont{J.}~\bibnamefont{Flouquet}},
  \bibinfo{journal}{Phys. Rev. Lett.} \textbf{\bibinfo{volume}{117}},
  \bibinfo{pages}{046401} (\bibinfo{year}{2016}).

\bibitem[{\citenamefont{Kotegawa et~al.}(2015)\citenamefont{Kotegawa, Fukumoto,
  Toyama, Tou, Harima, Harada, Kitaoka, Haga, Yamamoto, {\=O}nuki
  et~al.}}]{kotegawa201573ge}
\bibinfo{author}{\bibfnamefont{H.}~\bibnamefont{Kotegawa}},
  \bibinfo{author}{\bibfnamefont{K.}~\bibnamefont{Fukumoto}},
  \bibinfo{author}{\bibfnamefont{T.}~\bibnamefont{Toyama}},
  \bibinfo{author}{\bibfnamefont{H.}~\bibnamefont{Tou}},
  \bibinfo{author}{\bibfnamefont{H.}~\bibnamefont{Harima}},
  \bibinfo{author}{\bibfnamefont{A.}~\bibnamefont{Harada}},
  \bibinfo{author}{\bibfnamefont{Y.}~\bibnamefont{Kitaoka}},
  \bibinfo{author}{\bibfnamefont{Y.}~\bibnamefont{Haga}},
  \bibinfo{author}{\bibfnamefont{E.}~\bibnamefont{Yamamoto}},
  \bibinfo{author}{\bibfnamefont{Y.}~\bibnamefont{{\=O}nuki}},
  \bibnamefont{et~al.}, \bibinfo{journal}{J. Phys. Soc. Jpn.}
  \textbf{\bibinfo{volume}{84}}, \bibinfo{pages}{054710}
  (\bibinfo{year}{2015}).

\bibitem[{\citenamefont{Miyake et~al.}(2008)\citenamefont{Miyake, Aoki, and
  Flouquet}}]{doi:10.1143/JPSJ.77.094709}
\bibinfo{author}{\bibfnamefont{A.}~\bibnamefont{Miyake}},
  \bibinfo{author}{\bibfnamefont{D.}~\bibnamefont{Aoki}}, \bibnamefont{and}
  \bibinfo{author}{\bibfnamefont{J.}~\bibnamefont{Flouquet}},
  \bibinfo{journal}{J. Phys. Soc. Jpn.} \textbf{\bibinfo{volume}{77}},
  \bibinfo{pages}{094709} (\bibinfo{year}{2008}).

\bibitem[{\citenamefont{Posp{\'{i}}{\v{s}}il
  et~al.}(2017)\citenamefont{Posp{\'{i}}{\v{s}}il, Haga, Kambe, Tokunaga,
  Tateiwa, Aoki, Honda, Nakamura, Homma, Yamamoto et~al.}}]{PhysRevB.95.155138}
\bibinfo{author}{\bibfnamefont{J.}~\bibnamefont{Posp{\'{i}}{\v{s}}il}},
  \bibinfo{author}{\bibfnamefont{Y.}~\bibnamefont{Haga}},
  \bibinfo{author}{\bibfnamefont{S.}~\bibnamefont{Kambe}},
  \bibinfo{author}{\bibfnamefont{Y.}~\bibnamefont{Tokunaga}},
  \bibinfo{author}{\bibfnamefont{N.}~\bibnamefont{Tateiwa}},
  \bibinfo{author}{\bibfnamefont{D.}~\bibnamefont{Aoki}},
  \bibinfo{author}{\bibfnamefont{F.}~\bibnamefont{Honda}},
  \bibinfo{author}{\bibfnamefont{A.}~\bibnamefont{Nakamura}},
  \bibinfo{author}{\bibfnamefont{Y.}~\bibnamefont{Homma}},
  \bibinfo{author}{\bibfnamefont{E.}~\bibnamefont{Yamamoto}},
  \bibnamefont{et~al.}, \bibinfo{journal}{Phys. Rev. B}
  \textbf{\bibinfo{volume}{95}}, \bibinfo{pages}{155138}
  (\bibinfo{year}{2017}).

\bibitem[{\citenamefont{Sakakibara et~al.}(1994)\citenamefont{Sakakibara,
  Mitamura, Tayama, and Amitsuka}}]{sakakibara1994faraday}
\bibinfo{author}{\bibfnamefont{T.}~\bibnamefont{Sakakibara}},
  \bibinfo{author}{\bibfnamefont{H.}~\bibnamefont{Mitamura}},
  \bibinfo{author}{\bibfnamefont{T.}~\bibnamefont{Tayama}}, \bibnamefont{and}
  \bibinfo{author}{\bibfnamefont{H.}~\bibnamefont{Amitsuka}},
  \bibinfo{journal}{Jpn. J. Appl. Phys.} \textbf{\bibinfo{volume}{33}},
  \bibinfo{pages}{5067} (\bibinfo{year}{1994}).

\bibitem[{\citenamefont{Kittaka et~al.}(2014)\citenamefont{Kittaka, Kasahara,
  Sakakibara, Shibata, Yonezawa, Maeno, Tenya, and
  Machida}}]{PhysRevB.90.220502}
\bibinfo{author}{\bibfnamefont{S.}~\bibnamefont{Kittaka}},
  \bibinfo{author}{\bibfnamefont{A.}~\bibnamefont{Kasahara}},
  \bibinfo{author}{\bibfnamefont{T.}~\bibnamefont{Sakakibara}},
  \bibinfo{author}{\bibfnamefont{D.}~\bibnamefont{Shibata}},
  \bibinfo{author}{\bibfnamefont{S.}~\bibnamefont{Yonezawa}},
  \bibinfo{author}{\bibfnamefont{Y.}~\bibnamefont{Maeno}},
  \bibinfo{author}{\bibfnamefont{K.}~\bibnamefont{Tenya}}, \bibnamefont{and}
  \bibinfo{author}{\bibfnamefont{K.}~\bibnamefont{Machida}},
  \bibinfo{journal}{Phys. Rev. B} \textbf{\bibinfo{volume}{90}},
  \bibinfo{pages}{220502} (\bibinfo{year}{2014}), \bibinfo{note}{{S}upplemental
  {M}aterial}.

\bibitem[{not()}]{note}
\bibinfo{note}{From Maxwell equation $\mathrm{div}{\bm B}\!=\!0$, we expect
  $dH_x/dx\!=\! dH_y/dy\! =\! -\frac{1}{2}dH_z/dz$ (in vacuum) at the magnet
  center. The vertical field gradient $G\!=\!8$~T/m along the $b$ axis of the
  sample thus produce a field gradient of order 4~T/m along the $c$ axis. An
  off-center displacement of the sample of 1mm then results in a $c$-axis field
  of 4~mT.}

\bibitem[{\citenamefont{Taufour et~al.}(2016)\citenamefont{Taufour,
  Kaluarachchi, and Kogan}}]{PhysRevB.94.060410}
\bibinfo{author}{\bibfnamefont{V.}~\bibnamefont{Taufour}},
  \bibinfo{author}{\bibfnamefont{U.~S.} \bibnamefont{Kaluarachchi}},
  \bibnamefont{and} \bibinfo{author}{\bibfnamefont{V.~G.} \bibnamefont{Kogan}},
  \bibinfo{journal}{Phys. Rev. B} \textbf{\bibinfo{volume}{94}},
  \bibinfo{pages}{060410} (\bibinfo{year}{2016}).

\bibitem[{\citenamefont{Karube et~al.}(2012)\citenamefont{Karube, Hattori,
  Kitagawa, Ishida, Kimura, and Komatsubara}}]{PhysRevB.86.024428}
\bibinfo{author}{\bibfnamefont{K.}~\bibnamefont{Karube}},
  \bibinfo{author}{\bibfnamefont{T.}~\bibnamefont{Hattori}},
  \bibinfo{author}{\bibfnamefont{S.}~\bibnamefont{Kitagawa}},
  \bibinfo{author}{\bibfnamefont{K.}~\bibnamefont{Ishida}},
  \bibinfo{author}{\bibfnamefont{N.}~\bibnamefont{Kimura}}, \bibnamefont{and}
  \bibinfo{author}{\bibfnamefont{T.}~\bibnamefont{Komatsubara}},
  \bibinfo{journal}{Phys. Rev. B} \textbf{\bibinfo{volume}{86}},
  \bibinfo{pages}{024428} (\bibinfo{year}{2012}).

\bibitem[{\citenamefont{Mineev}(2015)}]{PhysRevB.91.014506}
\bibinfo{author}{\bibfnamefont{V.~P.} \bibnamefont{Mineev}},
  \bibinfo{journal}{Phys. Rev. B} \textbf{\bibinfo{volume}{91}},
  \bibinfo{pages}{014506} (\bibinfo{year}{2015}).

\end{thebibliography}

\end{document}